\documentclass[fleqn,usenatbib]{mnras}

\usepackage{amssymb}	

\usepackage{newtxtext,newtxmath, hyperref}
\usepackage{soul}
\usepackage[dvipsnames]{xcolor}
\colorlet{LightRubineRed}{RubineRed!70!}
\colorlet{color1}{green!10!orange!90!}

\usepackage[T1]{fontenc}
\usepackage{array}

\DeclareRobustCommand{\VAN}[3]{#2}
\let\VANthebibliography\thebibliography
\def\thebibliography{\DeclareRobustCommand{\VAN}[3]{##3}\VANthebibliography}

\usepackage{graphicx,subfigure}	
\usepackage{float}
\usepackage{amsmath}	

\usepackage[normalem]{ulem}

\usepackage{caption}

\title[Narrow-band giant pulses from the Crab pulsar]{Narrow-band giant pulses from the Crab pulsar}

\author[Thulasiram $\&$ Lin]{
Parasar Thulasiram$^{1}$\thanks{E-mail: p.thulasiram@mail.utoronto.ca} and 
Hsiu-Hsien Lin$^{1}$
\\
$^{1}$Canadian Institute for Theoretical Astrophysics, 60 Saint George St, Toronto, ON M5S 3H8, Canada
}

\date{Accepted XXX. Received YYY; in original form ZZZ}

\pubyear{2021}

\begin{document}
\label{firstpage}
\pagerange{\pageref{firstpage}--\pageref{lastpage}}
\maketitle

\begin{abstract}

We used a new spectral-fitting technique to identify a subpopulation of 6 narrow-band giant pulses from the Crab pulsar out of a total of 1578. These giant pulses were detected in 77 minutes of observations with the 46-m dish at the Algonquin Radio Observatory at 400--800 MHz. The narrow-band giant pulses consist of both main- and inter-pulses, thereby being more likely to be caused by an intrinsic emission mechanism as opposed to a propagation effect. Fast Radio Bursts (FRBs) have demonstrated similar narrow-band features while only little has been observed in the giant pulses of pulsars. We report the narrow-band giant pulses with $\Delta\nu$/$\nu$ on the order of 0.1, which is close to the value of 0.05 reported for the repeater FRB 20190711A. Hence, the connection between FRBs and giant pulses of pulsars is further established.
\end{abstract}

\section{Introduction}\label{section: introduction}

Giant pulses from the Crab pulsar have demonstrated fluences of thousands of Jy ms and burst energies up to 10$^{31}$ erg \citep{2004ApJ...612..375C, 2015ApJ...802..130H, 2016ApJ...833...47H, 2016JPlPh..82c6302E, 2019MNRAS.490L..12B, 2021arXiv210508851B}. They have shown various features in the spectrum, including broad-bandedness \citep{2012A&A...538A...7K, 2016ApJ...829...62E} and drifting \citep{2003Natur.422..141H, 2007ApJ...670..693H, 2021arXiv210508851B}, as well as multiple peaks in the temporal profile {\citep{2003Natur.422..141H, 2021arXiv210508851B}}. Main-pulses from the Crab at high frequencies (8--10.5 GHz) consist of micro-bursts, each of which can be resolved into overlapping strongly and randomly polarized nano-shots \citep{2007ApJ...670..693H}. Some speculate that a generic giant pulse from other pulsars such as PSR B1937+21 and PSR B1821$-$24A could follow a similar pattern \citep{2015ApJ...803...83B}.

Fast Radio Bursts (FRBs) are bright radio transients with durations of milliseconds, fluences from a few to hundreds of Jy ms, and origins from galactic to cosmological distances {\citep{2007Sci...318..777L,2013Sci...341...53T, 2015Natur.528..523M, 2019Natur.566..230C, 2020Natur.587...54C, 2020Natur.587...59B}}. Hundreds of FRBs were reported in the past decade {\citep{2016PASA...33...45P, 2021arXiv210604352T}}. While some FRBs demonstrated repetition \citep{2016Natur.531..202S,2019ApJ...885L..24C,2020ApJ...891L...6F} and {two} with periodic activity {\citep{2020Natur.582..351C, 2020MNRAS.495.3551R}}, the majority of FRBs appeared non-repeating. The repeating FRBs typically showed narrow-band features and downward frequency drifts across the dynamic spectrum \citep{2019Natur.566..235C,2019ApJ...876L..23H,2020ApJ...891L...6F}, and the non-repeating FRBs usually showed broad-band features in the dynamic spectrum \citep{2019Natur.566..230C, 2019ApJ...872L..19M}. Similar to the Crab, FRBs show a diversity in their temporal and spectral profiles \citep{2016PASA...33...45P,2019ApJ...877L..19G,2019Natur.566..230C,2019ApJ...885L..24C,2020ApJ...891L...6F, 2021arXiv210604356P}.

Recently, the repeater FRB 20190711A showed highly band-limited features, for which the spectral width is only 65$\pm$7 MHz at a central frequency of 1355$\pm$3 MHz \citep{2021MNRAS.500.2525K}. Moreover, PSR J0540$-$6919 recently showed similar narrow features \citep{2021arXiv210509096G}. As giant pulses have burst energies only a few orders of magnitude lower than FRBs, they have been associated with FRBs \citep{2016MNRAS.457..232C, 2016MNRAS.458L..19C}. The NASA Deep Space Network recently detected a burst from the repeating FRB 20200120E that shows a sub-100-nanosecond structure whose luminosity is $\sim$500 times brighter than the Crab pulsar's nano-shots \citep{2021arXiv210510987M, 2007ApJ...670..693H}. In addition, the Crab pulsar and a few repeating FRBs share similar power-law indices for their differential distributions of energy. The Crab, FRB 121102, and FRB 180916.J0158+65 power-law indices range from $2.1$ to $3.1$ \citep{2012ApJ...760...64M}, from $1.6$ to $1.8$ \citep{2019ApJ...882..108W}, and around $2.3$ \citep{2020Natur.582..351C}, respectively. Lastly, FRBs and giant pulses have similarities in hypothesized emission mechanisms that we will further discuss in section \ref{section 3.3: theory}. As such, we are motivated to study the connections between FRBs and giant pulses.

We report a narrow-band giant pulse subpopulation from the Crab pulsar in this paper, which further connects FRBs to pulsars' giant pulses. We structure the paper as follows: Section \ref{section: methods and analysis} will show the procedure of data pre-processing, illustrate a new technique to correct the dispersion measure (DM) through Singular Value Decomposition (SVD), and propose a new technique to classify a giant pulse as either broad- or narrow-band based on its spectral properties. Section \ref{section: discussion} will discuss the subpopulation's statistics, its properties, and its potential connections to FRBs. Section \ref{conclusion} will summarize the results and encourage future research avenues.

\section{Methods and Analysis}\label{section: methods and analysis}

\subsection{Data recording and giant-pulse search}\label{subsection: data recording and giant-pulse search}

We observed the Crab pulsar with the 46-m dish at the Algonquin Radio Observatory (ARO) on 2018 April 25 in two scans for a total duration of 77 minutes (49 min + 28 min; session 1 + session 2). We processed the baseband voltage data using a CHIME acquisition board \citep{2016JAI.....541004B}. This digitized the signals, passed them through a polyphase filter, and recorded them in the standard VLBI Data Interface Format (VDIF) \citep{2009evlb.confE..42W}. We channelized the data into 1024 frequency channels from 400 to 800 MHz for two linear polarizations with a time resolution of 2.56 $\mu$s.

The giant-pulse search algorithm coherently dedispersed the VDIF data with an initial DM value of $56.77$ pc cm$^{-3}$ \citep{1993MNRAS.265.1003L}\footnote{\url{http://www.jb.man.ac.uk/~pulsar/crab.html}}. After summing the power over all frequency channels and both polarization components, the algorithm searched for triggers with a minimal signal-to-noise ratio (S/N) of 5 in a $0.328$ ms rolling boxcar window. We describe a DM optimization procedure in section \ref{optimal_DM}. We re-searched the data with the optimized DM of 56.7563 pc cm$^{-3}$, which yielded 20085 triggers with {S/N > $5$}. We saved 12.8 ms of data for each trigger. We determined the triggers' rotational phase with TEMPO2 \citep{2006MNRAS.369..655H} using an ephemeris from Jodrell Bank \footnotemark[1] \citep{1993MNRAS.265.1003L}. We identified 1578 triggers with S/N > $9$ and the rotational phase being either main-pulse or inter-pulse as giant pulses for further investigation in this study.

\subsection{Optimal DM determination}\label{optimal_DM}
The initial DM systematically underestimated the S/N, thereby reducing some potential pulses' S/N below the detection threshold. It also changed dynamic spectra structure, limiting the potential for a spectral study. Hence, we corrected the DM. 
We propose to use Singular Value Decomposition (SVD), which has been applied to improve the precision of pulsar timing \citep{2018MNRAS.475.1323L}, to automatically optimize the DM for the giant pulses. SVD decomposes the dynamic spectrum of the giant pulse as the sum of individual modes of dynamic spectra, which can then be analyzed as a function of DM.

\begin{figure*}
\centering
\includegraphics[width=\textwidth]{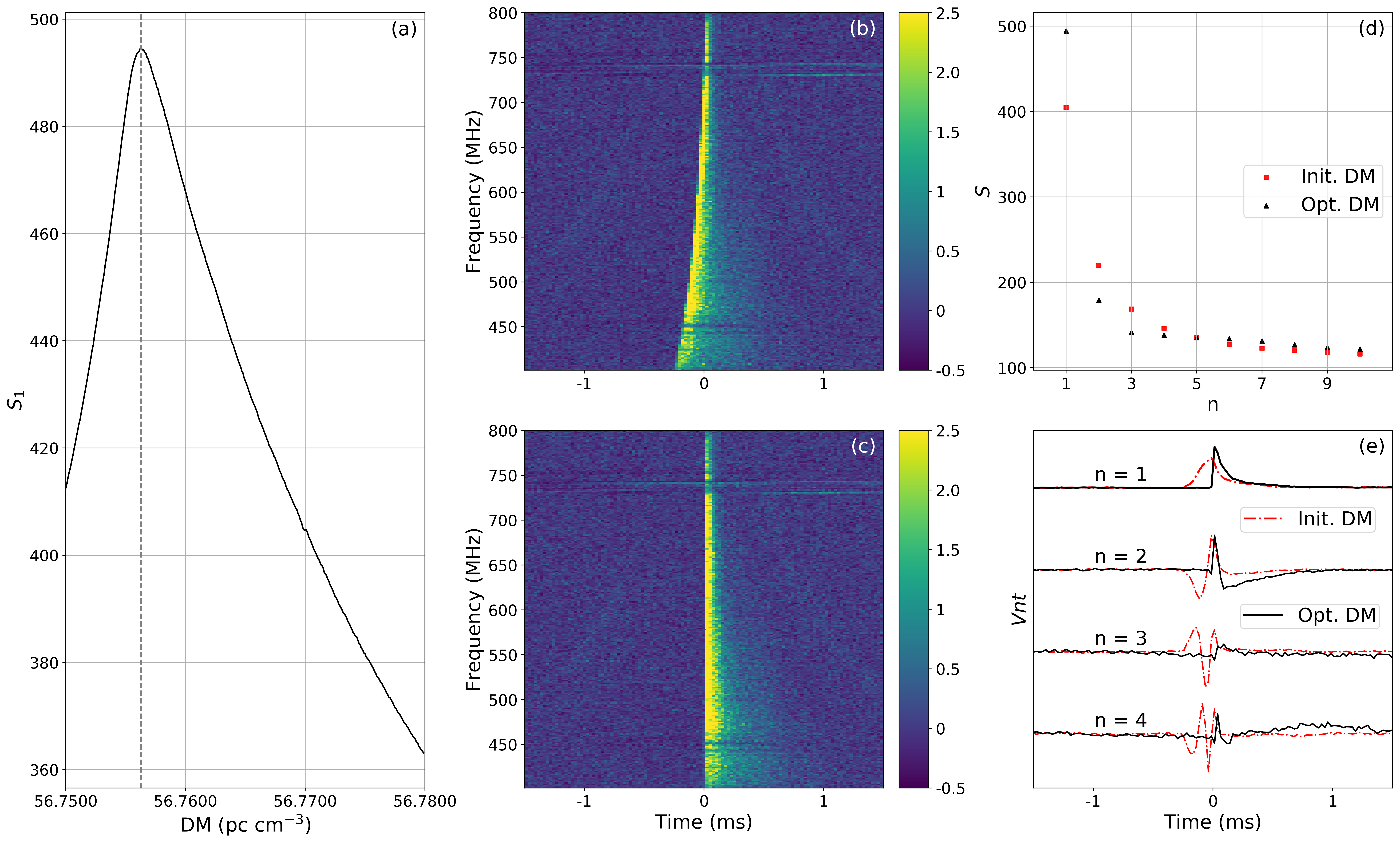}
\caption{\textbf{DM optimization with the SVD technique.} Panel $(a)$ shows the $S_1$ value as a function of DM. The optimal DM of 56.7563 pc cm$^{-3}$, which is marked with the dashed line, was determined by the corresponding maximal value of $S_1$. Panel $(b)$ shows the dynamic spectrum of the brightest broad-band pulse of session 2 with the initial DM of 56.77 pc cm$^{-3}$. Panel $(c)$ shows the dynamic spectrum with the optimal DM of 56.7563 pc cm$^{-3}$. Panel $(d)$ shows the comparison of $S$ between the SVD decomposition of the dynamic spectrum with the initial (red) and optimal (black) DM values, abbreviated init. and opt., respectively. Panel $(e)$ shows the the comparison of $V_{nt}$ between the SVD decomposition of the dynamic spectrum with the initial (red) and optimal (black) DM values.}
\label{figure: DMcorr}
\end{figure*}

Specifically, we decomposed the dynamic spectrum of a bright pulse into one eigenvalue and two {eigenfunctions}: 
\begin{equation}\label{equation: svd}
P_{ft} =\sum_n {U_{fn}}{S_{n}}{V_{nt}^{\top}},
\end{equation}
where $P_{ft}$ is the pulse's dynamic spectrum, and for each mode $n$: $U_{fn}$ is the eigenfunction in frequency $f$, $S_{n}$ is the eigenvalue, and $V_{nt}^{\top}$ represents the transpose eigenfunction in time. When we change the pulse's DM, the corresponding primary eigenvalue was also changed. We determined the dynamic spectrum's optimal DM value with the maximal primary eigenvalue. 

We applied the procedure to session 2's brightest broad-band pulse. We adjusted the dynamic spectrum by incoherent dedispersion within a range of 56.7550 to 56.7580 pc cm$^{-3}$ and with a step of 0.0001 pc cm$^{-3}$. We then decomposed each dynamic spectrum into the eigenvalue and the two eigenfunctions using SVD. Panel $a$ in Fig. \ref{figure: DMcorr} shows the primary eigenvalue across a range of DM values. We determined the optimal DM value, 56.7563 pc cm$^{-3}$, by the corresponding maximal primary eigenvalue, and we re-searched the full data using this DM. Panels $b$ and $c$ in Fig. \ref{figure: DMcorr} show the pulse's dynamic spectrum with the initial (56.77 pc cm$^{-3}$) and optimal DM values, respectively. Panels $d$ and $e$ in Fig. \ref{figure: DMcorr} show the pulse's eigenvalue and eigenfunction in time across modes with the initial and optimal DM values, respectively. The eigenfunction in time's primary mode represents the frequency-averaged pulse profile, for which the optimal DM profile has a higher amplitude compared to the initial DM profile. Furthermore, the optimal DM profile shows the noise-level in the third mode, while the initial DM profile does not show the noise-level until beyond the fourth mode.

\subsection{Gain correction}\label{subsection: gain correction}
We calibrated the frequency-dependent gains, which are closely related to the observed spectral structure. To do this, we divided each time-slice of the dynamic spectrum by the recording session's average gain profile. We determined the profiles by noise-subtracting each session's 1000 brightest pulses, aligning them in time, and then averaging over time in each polarization. \citet{2021arXiv210508851B} contains more details.

\subsection{Flux Calibration}
Flux from the Crab nebula dominated the off-pulse emission. We subtracted the Crab's average off-pulse emission by B0355+54's average off-pulse emission to measure the flux contribution from the Crab nebula. B0355+54 was another pulsar recorded in the same run \citep{2021arXiv210508851B}. The giant pulse's flux $S_{\text{GP}}$ is then \begin{equation}
    S_{\text{GP}}(f)= S_{\text{nebula}}(f) \times I_{\text{GP}}(f) / I_{\text{nebula}}(f), 
\end{equation}
where $S_{\text{nebula}}(f) \simeq 955 $ Jy ($f$/1 GHz)$^{-0.27}$ is the Crab nebula's nebular flux spectrum, $I_{\text{GP}}$ is the giant pulse's intensity, and $I_{\text{nebula}}$ is the {nebula's} intensity. \citet{2021arXiv210508851B} contains more details.

\subsection{Radio-frequency interference (RFI) masking}\label{subsection: RFI masking}
For each giant pulse, we have 12.8 ms of data. We defined the on-pulse region as
$-0.256$ to 1.28 ms, where the peak-value of time-series is centered at 0 ms. We chose this range to include precursor-like or strong-scattering features. If the peak-value was located near the time-window's boundary, the boundary, instead, was used as a cutoff. We defined the remaining 11.264 ms as the off-pulse region.

We identified and masked the RFI channels through the following schema. First, for each frequency channel, we calculated the mean and standard deviation over the off-pulse time bins, yielding two functions of frequency $m_{\text{off}} $ and $ \sigma_{\text{off}} $. We then calculated the gradients of these functions with respect to frequency, $\nabla m_{\text{off}} $ and $ \nabla \sigma_{\text{off}}$.
We determined the means and standard deviations of
$\nabla m_{\text{off}}, \sigma_{\text{off}},$ and $\nabla \sigma_{\text{off}}$. 
Finally, we masked channels with $\nabla m_{\text{off}}, \sigma_{\text{off}},$ or $\nabla \sigma_{\text{off}}$ exceeding {$\pm$ 1 standard deviation of their respective frequency-means}. 

Certain channels demonstrated large negative dips in a frequency sub-band due to noise-subtraction and packet loss from coherent dedispersion. We analyzed these pulses separately. 

After masking the RFI channels, we averaged the dynamic spectrum over the on-pulse region of 1.536 ms. We rebinned the averaged dynamic spectrum from 1024 to 64 frequency channels to determine the pulse's intensity spectrum.
We further masked rebinned channels that consist of more than 40 \% by RFI channels.

\subsection{Spectral fitting}\label{subsection: spectral fitting}
Traditionally, a Crab giant pulse's intensity spectrum is fit with a power-law \citep{2010A&A...515A..36K, 2012A&A...538A...7K, 2016ApJ...829...62E}. For broad-band pulses, this is a strong fit \citep{2018IAUS..337..378M}; however, narrow-band pulses display nonlinear structure in amplitude versus frequency in log-log space and thus require a fit with more curvature.

\begin{figure*}
\centering     
\subfigure[$\beta$ v.s. $\alpha$]{\label{fig:stat_beta_alpha}\includegraphics[width=0.33\textwidth]{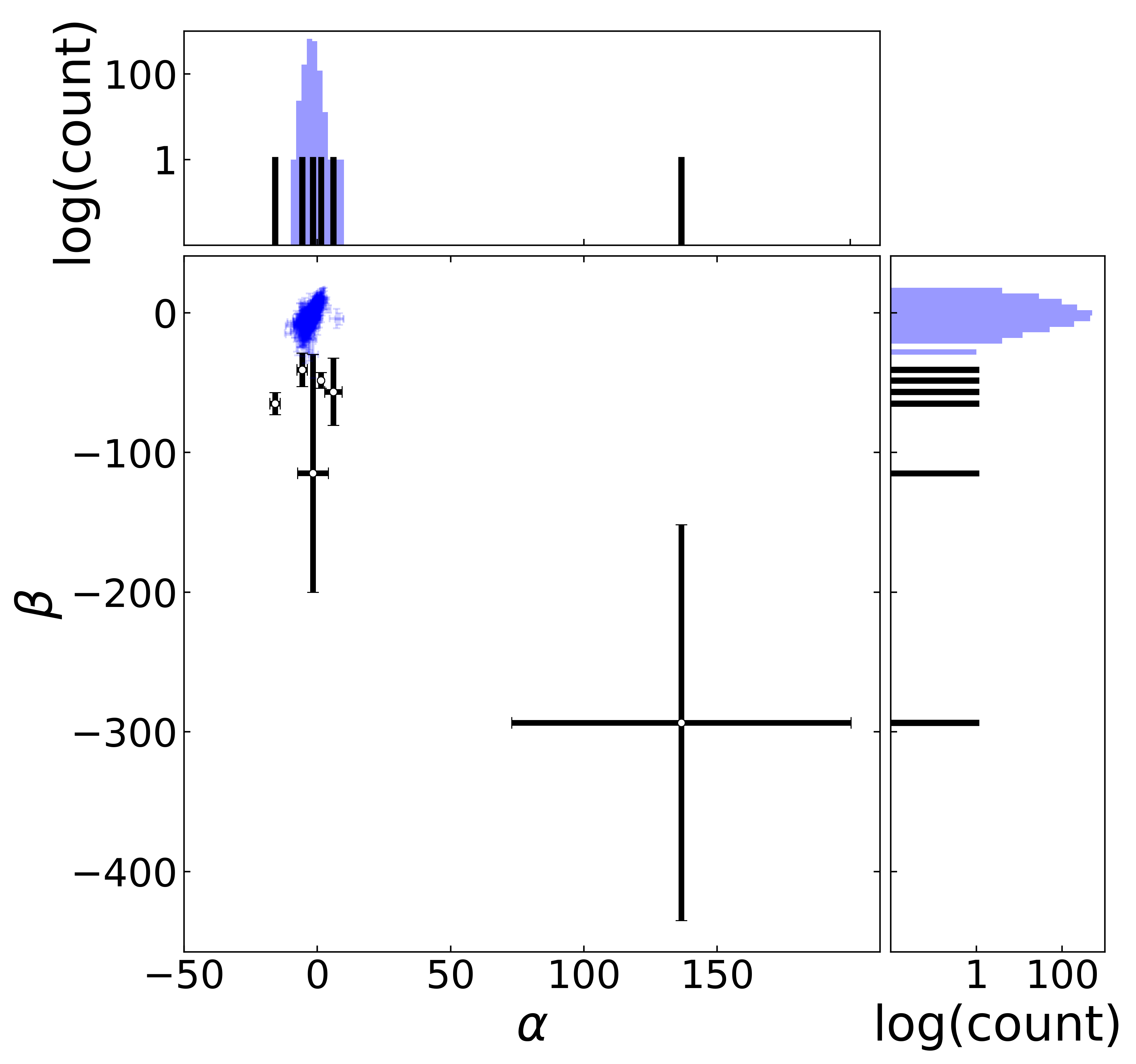}}
\subfigure[$\kappa$ v.s. $\alpha$]{\label{fig:stat_kappa_alpha}\includegraphics[width=0.32\textwidth]{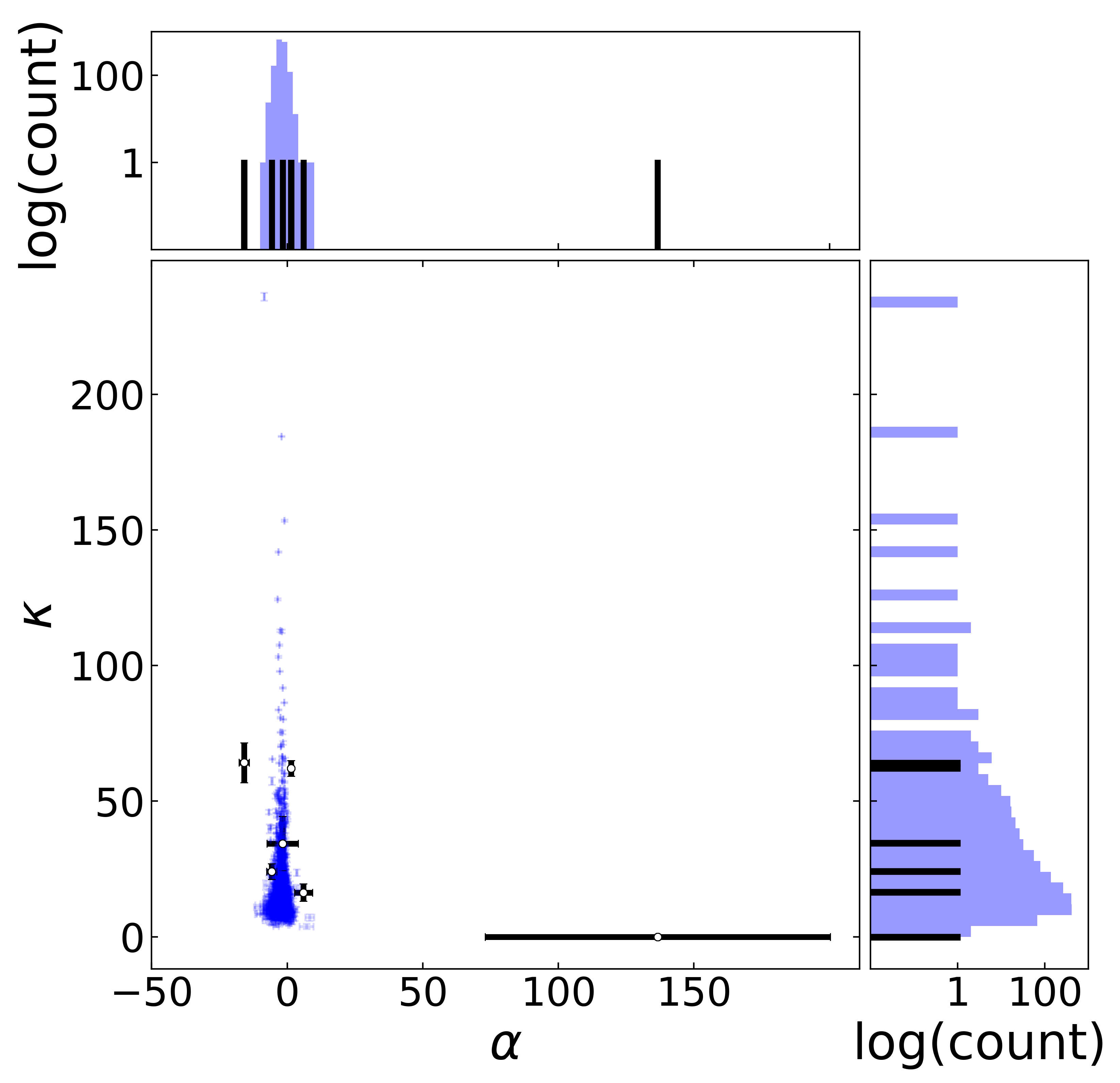}}
\subfigure[$\kappa$ v.s. $\beta$]{\label{fig:stat_kappa_beta}\includegraphics[width=0.32\textwidth]{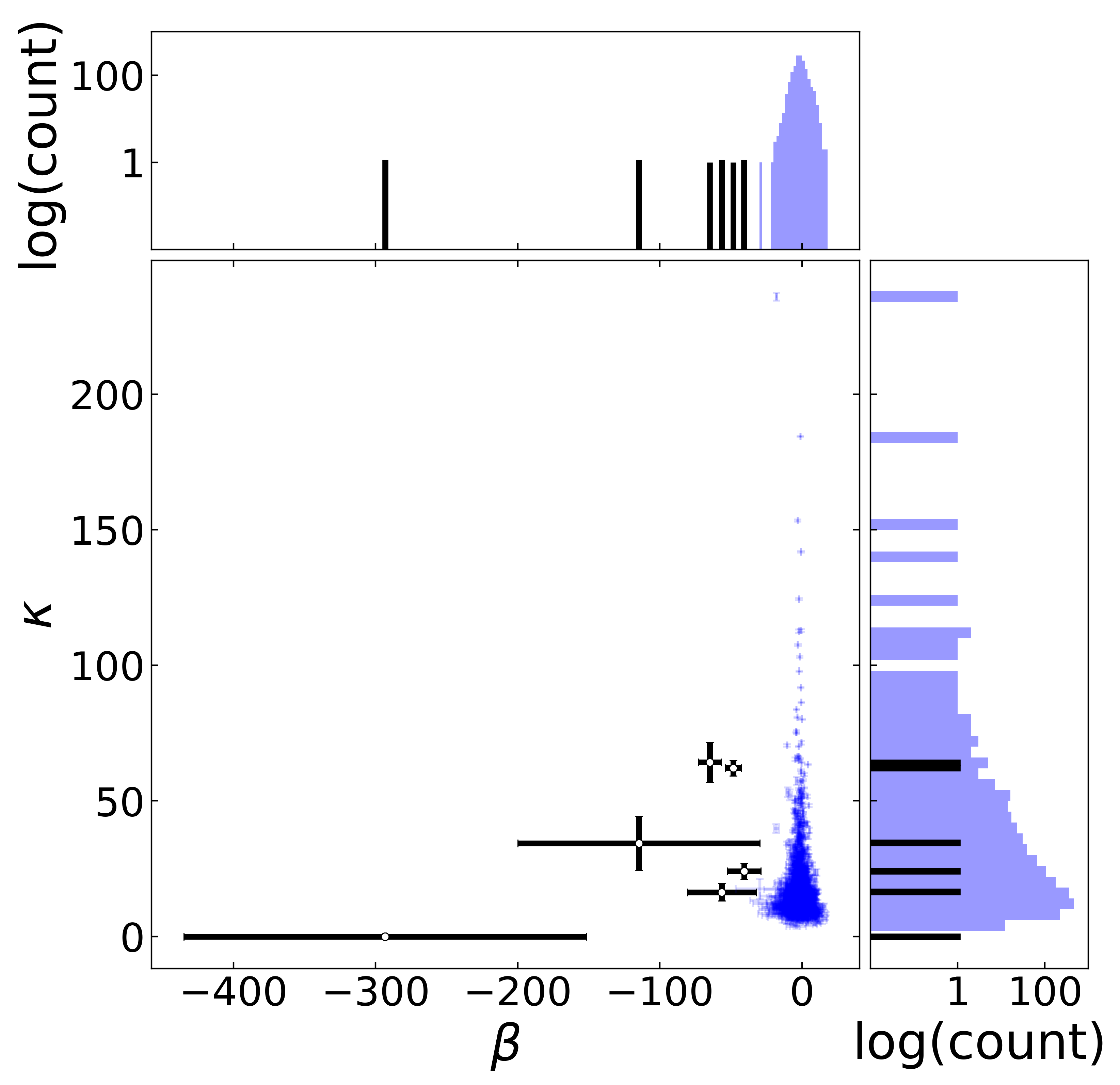}}
\caption{\textbf{The distributions of the fitting parameters $\alpha$, $\beta$, and $\kappa$.} In each of the middle panels, the points with blue errorbars represent broad-band pulses (1572 of them) while the points with black errorbars represent narrow-band pulses (6 of them). The errorbars represent the 68$\%$ confidence interval of the fitted result. Furthermore, for each of the following sub-plots, the middle panel shows the scatter plot of the $(x,y)$ data, the top panel is the histogram of the $x$-data, and the right panel is the histogram of the $y$-data.  Panel (a): $\beta$ vs $\alpha$ . A clear separation of the narrow- and broad-band populations suggests $\beta$ is a discriminating parameter between them. Panel (b): $\kappa$ vs $\alpha$. A lack of separation of the populations in the middle and top panel suggests $\alpha$ is not a discriminating parameter. Panel (c): $\kappa$ vs $\beta$. The separation of the populations suggests $\beta$ is a discriminating parameter, and the separation of narrow-band pulses in the right panel suggests they occur at a wide range of brightnesses. As discussed in section \ref{subsection: spectral fitting}, $\kappa$ is coupled to $\alpha;$ therefore, the narrow-band $\kappa$ values are not fully accurate in panels (b) and (c), but the fact that narrow-band pulses are observed at a wide range of brightnesses remains even after refitting.}
\label{figure: statistical}
\end{figure*}

Hence, we applied the following fitting function to the 1578 pulses using least-squares optimization:
\begin{equation}\label{eq:spectral}
    I = \kappa \left( \frac{f}{f_0} \right )^{\alpha + \beta \log \left ( \frac{f}{f_0} \right ) }, 
\end{equation}
where $f$ is a frequency value between 400 and 800 MHz, $I$ is the amplitude (arbitrary units) at a given frequency, $\alpha, \beta, $ and $ \kappa$ are the free parameters to be fit and $f_0 = 566$ MHz is the reference frequency, as it is the log-scale centre. To avoid over or underestimating the model in the log-log space, we applied an upper limit of $10^3$ and a lower limit of $10^{-1}$ for the amplitude of the spectrum. We estimated the fitted parameters' uncertainty in eq. \ref{eq:spectral} through the standard error propagation. Figure \ref{figure: statistical} shows the fitting results, and we will discuss the distribution of parameters in section \ref{subsection: a population of narrow-band pulses}.

We refer to Equation \ref{eq:spectral} as the quadratic fit in accordance with its intensity-frequency structure in log-log space. $\kappa$ represents the amplitude, $\alpha$ relates to the bandwidth center of the pulse relative to $f_0$, and $\beta$ relates to the bandwidth of the pulse. The parameters are coupled and can therefore result in poor fits if the reference frequency is too different from the narrow-band spectral centre. To correct this, we refitted the identified narrow-band pulses, which will be described in section \ref{subsection: a population of narrow-band pulses}, with reference frequencies much closer to the spectral centre. We display the resulting properties in Table \ref{table: table_narrow} with error propagation considering covariances. The coupling is not an issue for broad-band pulses.

\begin{figure*}
\centering
\includegraphics[width=\textwidth]{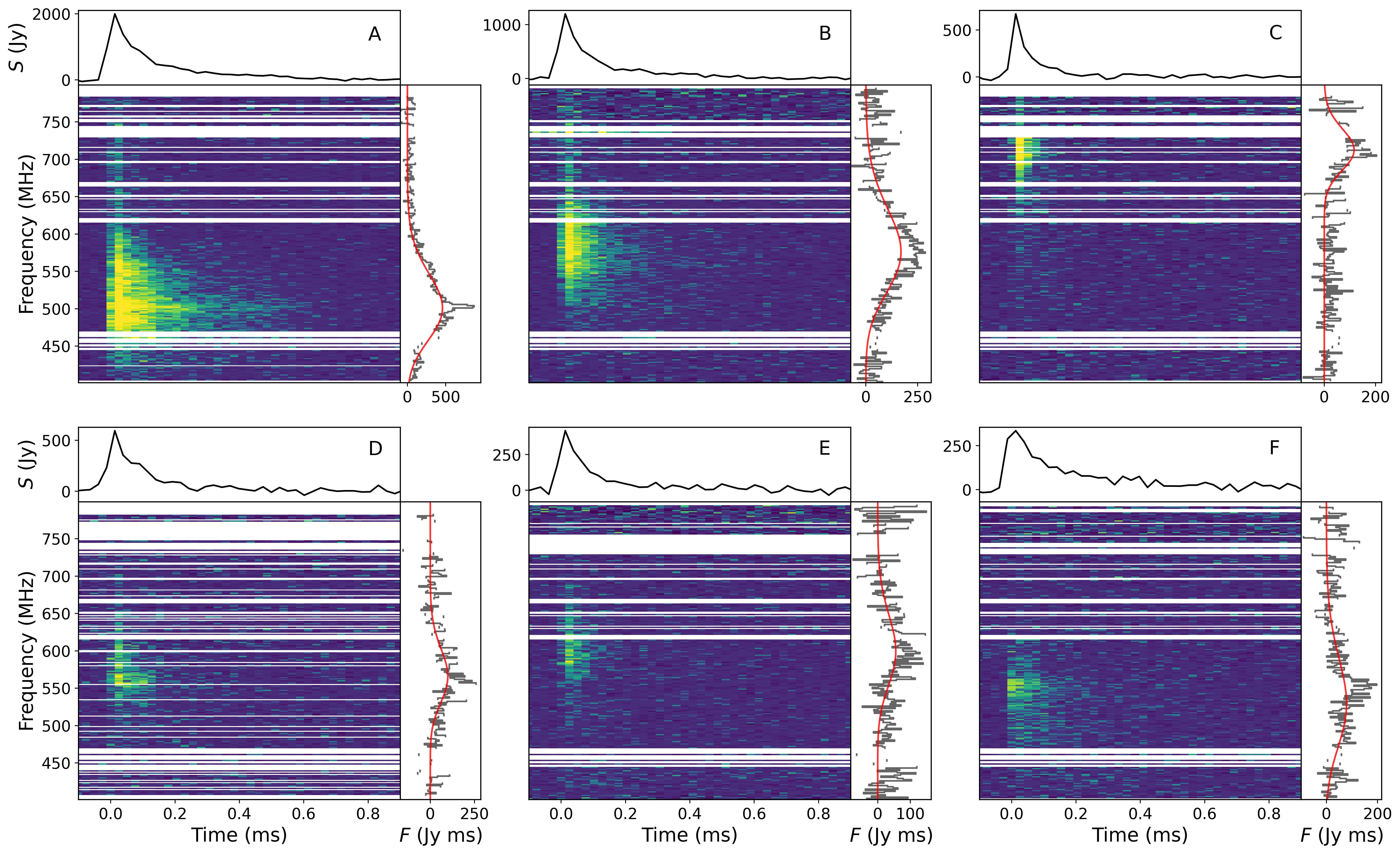}
\caption{{\bf The dynamic spectrum, time series, and spectral profile of the six narrow-band giant pulses.} Each central panel shows the frequency-time dynamic spectrum with a frequency resolution of 1.5625 MHz and a timing resolution of 25.6 $\mu$s, where the RFI channels are masked out. The upper panel shows the time series of the flux ($S$) in units of Jy, which is the frequency-averaged profile. The side panel shows the spectral profile  of the fluence ($F$) in units of Jy ms and the fitting model via the grey and red lines, respectively.}
\label{figure: gallery}
\end{figure*}

\section{Discussion}\label{section: discussion}

\subsection{A population of narrow-band pulses}\label{subsection: a population of narrow-band pulses}

In total, we identified six narrow-band giant pulses with three main-pulses and three inter-pulses, which represent a 0.4\% population in the 1578-pulse sample. Fig. \ref{figure: statistical} shows the distributions of the fitting parameters for the broad-band (in blue) and narrow-band (in black) pulse populations. Therefore, the clustering of the broad-band pulses, with narrow-band outliers in panels $a$ and $c$, distinguish the two populations.

Due to the high spectral curvature, narrow-band pulses show $\beta \leq -30$ in the $\beta$ value histograms in panels ($a$) and ($c$) of Fig. \ref{figure: statistical}. We further validated this threshold by inspecting the quadratic fit with $\alpha$ and $\kappa$ fixed at their population medians and noting that narrow-band features are strongly visible around $\beta \leq -30$. This indicates the method could be used to identify narrow-band pulses in a data-set. Furthermore, since the narrow-band and broad-band points overlap in panels ($a$) and ($b$), $\alpha$ is not a strong discriminator of narrow- vs broad-bandedness. We note the largest $\alpha$ in panels ($a$) and ($b$) of Fig. \ref{figure: statistical} corresponds to Pulse C in Fig. \ref{figure: gallery}. In general, a large $\alpha$ indicates a significant difference between the spectral center and the reference frequency.

We present a gallery of the flux-calibrated narrow-band pulses in Fig. \ref{figure: gallery} in descending brightness from A - F. We show the on-pulse-averaged spectral profile overlaid with the quadratic fitting in the right panel and the time-series in the top panel.
We display the properties of the narrow-band pulses in Table \ref{table: table_narrow}. We measured the peak flux and fluence using the data with a 25.6 $\mu$s time resolution and a 1.5625 MHz frequency resolution. The spectral centre $\nu$ refers to the frequency channel with the model's peak value while the spectral bandwidth $\Delta\nu$ refers to the full width at half maximum (FWHM). This table shows that narrow-bandedness is independent of pulse-brightness, as there are narrow-band pulses observed at a range of S/N, peak flux densities, and fluences. Furthermore, $\Delta \nu / \nu$ is independent of brightness but further data would be helpful to understand its distribution.

\begin{table}
\begin{tabular}{m{4.5em}m{2.6em}m{2.6em}m{2.6em}m{2.6em}m{2.6em}m{2.6em}}
\hline \hline
Pulse & A & B & C & D & E & F \\
\hline
S/N &57&32&16&13&11&9 \\
Type &MP&IP&MP&IP&MP&IP\\
$S_{\text{max}}$ (Jy) &1995(38)&1198(56)&669(29)&593(22)&419(48)&335(23) \\
$F$ (Jy ms) &285(1)&146(1)&52(1)&69(1)&50(1)&71(1) \\
$\nu$ (MHz) &500.9(3)&575.3(4)&713(1)&563(1)&602(2)&528(2) \\
$\Delta \nu$ (MHz) &100.8(5)&122.8(7)&71(3)&90.(2)&68(3)&133(4) \\
${\Delta \nu}$/${\nu}$ &0.201(1)&0.213(1)&0.100(4)&0.160(4)&0.112(5)&0.251(8) \\
\hline
\end{tabular}
\caption{\textbf{Properties of six narrow-band pulses.} The label of the pulse corresponds to Fig. \ref{figure: gallery}. The S/N is integrated over the whole 400 MHz bandwidth. MP and IP refer to main- and inter-pulse, respectively. $S_{\text{max}}$ refers to the peak flux density and $F$ to the fluence. The $\nu$ and $\Delta\nu$ represent the measured spectral center and bandwidth at FWHM, respectively. The statistical uncertainties represent the 68$\%$ confidence interval of the measurements.
}
\label{table: table_narrow}
\end{table}

\subsection{The microstructure in the brightest narrow-band pulse}\label{subsection: the microstructure in the brightest narrow-band pulse}

\citet{2018MNRAS.478.1209F} reported FRB 170827 with fine spectral modulations on a frequency scale of 1.5 MHz and striations on the scale of 100--200 KHz. It also displayed spiky emission features that can exceed 1 kJy in brightness in the frequency subband of 841--843 MHz. Motivated by this result, we further studied Pulse A's observably hyper-narrow-band intensity around 496--512 MHz. 

Pulse A shows a few bright and unresolved spots in the dynamic spectrum when we viewed it with the highest resolution in time (2.56 $\mu$s) and frequency (0.390625 MHz). These spots are referred to as the microstructure. The bright microstructures are around 496--512 MHz and 0--30 $\mu$s in the dynamic spectrum. 

Interstellar scintillation (ISS) is unlikely to cause the feature. The angular broadening of the Crab in very long baseline interferometry yields a scattering angle of $\theta \sim 1.5$ mas at 600 MHz for ISS \citep{2016ARep...60..211R}. Assuming thin-screen scattering with a scattering screen roughly {halfway} to the pulsar \citep{1976ApJ...209..578V}, the decorrelation bandwidth is $30$ kHz. This is much smaller than the $390$ KHz frequency resolution, so the scintillation is unlikely to cause the {microstructure}.

Through the following statistical analysis, we argue that the bright microstructures show no significance greater than 3$\sigma$. We first standardized the off-pulse region to obtain the mean and standard deviation as equal to 0 and 1, respectively. We divided the dynamic spectrum, which was pre-processed in section \ref{subsection: RFI masking}, by the standard deviation of the off-pulse for each frequency channel. This is the so-called on/off ratio.

We then compared the signal baseline between the on- and off-pulse regions for comparing the on-pulse bright microstructure to the noise. We found that the on-pulse region's signal baseline is 15 times higher than the off-pulse region's baseline. To estimate the significance of the on-pulse bright microstructures, we multiplied the off-pulse region by a factor of 15 (so-called the amplified off-pulse region) and used it as the noise-fluctuation for the on-pulse region. We then generated the cumulative distribution function (CDF) of the amplified region. We applied this process to each of the six {narrow-band} giant pulses.

Lastly, we determined the significance of the on-pulse bright microstructures with the amplified off-pulse region as the baseline. Considering the look-elsewhere effect for six narrow-band giant pulses, the z-score of 3 gives the p-value of 0.000225. This corresponds to the on/off ratio of 94 for the amplified off-pulse region CDF. We found that the on/off ratio of the bright microstructures has a range from 35 to 47, which is below the 3$\sigma$ significance.

\subsection{Possible physical scenarios of narrow-band emission}\label{section 3.3: theory}

A number of models predict {narrow-band} emission in radio transients. 
\citet{2017ApJ...842...35C} show that caustics produced by plasma lensing can strongly magnify ($\lesssim 10^2$) FRB signals on short timescales which look like narrow-band spectral peaks (0.1--1 GHz). These images can also interfere, creating even narrower ($\sim$ 1--100 MHz) frequency structures. If plasma lensing were the cause of narrow-band spectral structure, adjacent pulses in time should also demonstrate it, but we only observed 6 narrow-band pulses in a sample of 1578 at sporadic times, so this hypothesis is less likely. However, \citet{2021arXiv210508851B} argues that narrow-band features could arise from the interference of nano-shots that are nearly equally spaced, and these nano-shots could be echoes produced by plasma lensing in the pulsar wind. Therefore, plasma lensing could, in principle, cause the narrow-band features, though the predicted bandwidth of the narrow-band emission is still unknown.

\citet{2019MNRAS.485.4091M} argues that FRB emission can be modelled as synchrotron maser emission from decelerating
relativistic blast waves. This model explains FRBs' observed downward drifting \citep{2019Natur.566..235C,2019ApJ...876L..23H,2020ApJ...891L...6F} as arising from the blast-wave's deceleration coupled with induced Compton scattering at low frequencies and maser emission fall-off at high frequencies. For millisecond bursts, they predict a $\Delta \nu / \nu$ value near 1, which is above what we observe. However, they argue that this could be due to a simplistic treatment of low-frequency suppression of the spectral energy distribution due to Compton scattering and that a more accurate treatment of the suppression would lead to a sharper spectral peak. Thus, this model should not be ruled out as a potential explanation for narrow-band pulses.

Lastly, some models rely on the Lorentz factor of the particles that generate the radio waves. \citet{2021arXiv210207010L} proposes a model in which radio waves are
emitted by the Compton scattering of Alvfén waves interacting with a particle beam generated through reconnection events. They deduce a particle produces a set of emission bands with spectral separation $4 \gamma_0^2 k_w c$ where $\gamma_0$ is the particle beam's Lorentz factor and $k_w$ relates to the environmental turbulence properties. \citet{2020MNRAS.498.1397L} considers a model in which Alfvén waves carry released energy near the polar regions of the neutron star. When arriving at the charge starvation radius $R$, the waves generate an electric field that accelerates clumps of particles formed from two-stream instability. They argue the spectrum of radio waves from coherent curvature emission by the clumps can only be slightly broadened $(\Delta \nu/\nu \simeq (\gamma \ell_{\perp}/R)^2 \sim 0.1)$ due to the Doppler effect, and the spectrum can display variations over a narrow band from delays in the radiation arrival time. Here, $\gamma$ is the particle's Lorentz factor and $\ell_{\perp}$ is the particle clump's transverse size.

\subsection{Spectral similarities between pulsars and FRB 20190711A}

\citet{2020Natur.581..391M} reported FRB 20190711A with a bandwidth of $\sim$150 MHz at the \textit{L}-band, and later \citet{2021MNRAS.500.2525K} reported a repeating and narrow-band pulse from the same source with spectral centre $1355 \pm 3$ MHz and spectral bandwidth $65 \pm 7$ MHz, which corresponds to a $\Delta \nu / \nu$ value of $0.048 \pm 0.005$.

Recently, \citet{2021arXiv210509096G} observed "flux knots" in the giant pulses of PSR J0540$-$6919, which are single, bright, band-limited features that are symmetric in frequency response. They observed 24 flux knots in their three recording epochs totalling 5.7 hrs, with 18 showing FWHM less than 236 MHz when fit with a Gaussian profile. The mean FWHM of the 24 knots is 198 MHz and the mean FWHM of the 18 narrowest knots is 146 MHz. The narrowest band-limited pulse has FWHM $68.3 \pm 4.7$ MHz with a central frequency of $1472 \pm 2$ MHz, which corresponds to a $\Delta \nu / \nu$ value of roughly $0.046 \pm 0.003$ using the same definition as before. They determine the flux knots are unlikely to be caused by scintillation or lensing at the shock boundary.
 
 The $\Delta \nu / \nu$ values stated above are roughly a factor of two smaller than 6 narrow-band pulses that are listed in Table \ref{table: table_narrow}. This could potentially be explained by a point made in \citet{2021MNRAS.500.2525K}, that there could be emission below their detection threshold, leading to a smaller measured spectral bandwidth than was physically generated. This could be due to the far distance of FRB 20190711A ($z = 0.522$) \citep{2020Natur.581..391M,2021MNRAS.500.2525K}. Therefore, the $\Delta \nu / \nu$ values determined here are likely closer to that of FRB 20190711A than is immediately apparent. Moreover, by this argument, the $\Delta \nu / \nu$ value measured for the narrowest pulse of PSR J0540$-$6919 is likely due to intrinsic factors given its galactic origin. As the narrow-band pulses observed by \citet{2021arXiv210509096G} occupy a wider distribution of $\Delta \nu/ \nu$ above a threshold of $0.05,$ it is comparable to the pulses observed here and to FRB 20190711A.

In any case, the fact that FRB 20190711A and two pulsars show both broad-band and narrow-band features further connects FRBs and giant pulses from pulsars.

\section{Conclusions}\label{conclusion}
This paper reports a new narrow-band giant pulse subpopulation from the Crab pulsar identified between 400-800 MHz via a new fitting method. The approximate occurrence rate of these pulses is less than five in every thousand giant pulses. We observed the pulses with fluxes of several hundred to thousands of Jy and consisting of both main- and inter-pulses. Due to shared phenomenology, this discovery further connects giant pulses to FRBs.

\citet{2019ApJ...877L..19G} and \citet{2021MNRAS.500.2525K} argued that conducting sub-band burst search strategies 
can improve further research in this area. We encourage re-searching the existing giant-pulse data-sets with sub-band searching \citep{2012ApJ...760...64M,2016ApJ...829...62E,2018IAUS..337..378M,2020A&A...634A...3V}. Searching sub-bands would increase the chance that lower S/N narrow-band pulses with smaller $\Delta\nu$/$\nu$ would be found, especially in broad-band studies and studies of objects without a previously known DM \citep{2019ApJ...877L..19G}. Modifications in the definition of S/N could also be made to incorporate narrow-bandedness \citep{2019ApJ...877L..19G}.

Furthermore, while we found that the microstructure in the narrow-band giant pulses does not exceed 3$\sigma$ statistical significance, {others have investigated} the time and frequency microstructure in FRBs \citep{2018MNRAS.478.1209F, 2019ApJ...872L..19M, 2021NatAs.tmp...53N}. We encourage searching for this in giant pulses as exceptionally narrow-band features could reach the brightnesses observed in FRBs. For this reason, we encourage observing giant pulses at high time and frequency resolution \citep{2002MNRAS.334..523K, 2010ApJ...722.1908C, 2016ApJ...829...62E, 2020A&A...634A...3V, 2021arXiv210508851B} for studying the microstructure.

\section*{Acknowledgements}
We thank Ue-Li Pen and Marten H. van Kerkwijk as well as Akanksha Bij and Rebecca Lin who greatly aided in completing this project. Parasar Thulasiram thanks the 2020 University of Toronto Summer Undergraduate Research Program (SURP) in astronomy $\&$ astrophysics and the work-study program at the University of Toronto. We thank all members of the scintillometry group for their very helpful discussions. We thank Ziggy Pleunis for the discussion at CASCA 2021 AGM. We acknowledge the support of the Natural Sciences and Engineering Research Council of Canada (NSERC) (funding reference number RGPIN-2019-067, CRD 523638-201). We receive support from the Ontario Research Fund - Research Excellence Program (ORF-RE), the Canadian Institute for Advanced Research (CIFAR), the Canadian Foundation for Innovation (CFI), the Simons Foundation, Thoth Technology Inc., and the Alexander von Humboldt Foundation. Computations were performed on the Niagara and HPSS supercomputers at the SciNet HPC Consortium \citep{2010JPhCS.256a2026L,2019arXiv190713600P}. SciNet is funded by: the Canada Foundation for Innovation; the Government of Ontario; the Ontario Research Fund - Research Excellence; and the University of Toronto.

\section*{Data Availability}
The data analyzed in this study will be shared on reasonable request to the corresponding author.

\bibliographystyle{mnras}
\bibliography{references} 





\bsp	
\label{lastpage}
\end{document}